\documentclass[conference]{IEEEtran}
\IEEEoverridecommandlockouts
\usepackage{cite}
\usepackage{amsmath,amssymb,amsfonts}
\usepackage{algorithmic}
\usepackage{graphicx}
\usepackage{textcomp}
\usepackage{xcolor}
\usepackage{framed}
\usepackage{subcaption}
\colorlet{shadecolor}{yellow!20}
\usepackage[linesnumbered,ruled,noend]{algorithm2e}
\usepackage{tabularray}
\usepackage[inkscapeformat=png]{svg}
\usepackage[colorinlistoftodos]{todonotes}
\usepackage[export]{adjustbox}
\usepackage{soul}
\usepackage{mdframed}
\newmdenv[backgroundcolor=yellow!50, linewidth=0pt]{myframe}

\def\BibTeX{{\rm B\kern-.05em{\sc i\kern-.025em b}\kern-.08em
    T\kern-.1667em\lower.7ex\hbox{E}\kern-.125emX}}

\begin{document}
{


\twocolumn
\newtheorem{definition}{Definition}[section]

\title{Inferring Message Flows From System Communication Traces
}

\author{\IEEEauthorblockN{Bardia Nadimi, Hao Zheng}
\IEEEauthorblockA{\textit{dept. Computer Science and Engineering} \\
\textit{University of South Florida}\\
Tampa, United States \\
\{bnadimi, haozheng\}@usf.edu}
}

\maketitle
\vspace*{-15pt}

\begin{abstract}

This paper proposes a novel method for automatically inferring message flow specifications from the communication traces of a system-on-chip (SoC) design that captures messages exchanged among the components during a system execution. 
The inferred message flows characterize the communication and coordination of components in a system design for realizing various system functions, and they are essential for SoC validation and debugging.
The proposed method relieves the burden of manual development and maintenance of such specifications on human designers. 
Our method also uses a new accuracy metric, \emph{acceptance ratio}, to evaluate the quality of the mined specifications instead of the specification size often used in the previous work, enabling more accurate specifications to be mined. Furthermore, this paper introduces the concept of essential causalities to enhance the accuracy of the message flow mining and accelerate the mining process.
The effectiveness of the proposed method is evaluated on both synthetic traces and traces generated from executing several system models in GEM5.
In both cases, the proposed method achieves superior accuracies compared to a previous approach. 
Additionally, this paper includes some practical use cases.

\end{abstract}


\vspace*{-5pt}
\section{Introduction}
Modern System-on-Chip (SoC) designs consist of numerous functional blocks, each handling a distinct task and communicating through advanced protocols to enhance system functionality. These blocks operate concurrently, leading to simultaneous or interleaved system transactions. However, this concurrency and complex protocol interaction often result in runtime errors during debugging phases. An example of a standard SoC design, including components like CPUs, caches, and Network-on-Chip interconnects, is depicted in Fig.~\ref{fig:soc-ex}.

Having a precise, efficient, and comprehensive specification model is crucial for the validation, verification, and debugging of system designs.
Creating such a model for a system design can lead to the ability to simulate the system behavior and characteristics to foresee any design defects prior to fabricating the design. However, such models in practice are usually incomplete, inconsistent, ambiguous, or may even include errors. 
Therefore, the lack of such comprehensive models prevents the systematic analysis and validation of complex system designs. 


To address the above challenges, this paper introduces a new method for automatically inferring accurate message flow specifications from system communication traces.
The proposed method constructs a causality graph from input traces to understand message relationships.
Then, it selects a subset of message sequences in the causality graph as a potential specification model, which is evaluated on the input traces.
If the chosen model's accuracy is unsatisfactory, it will be refined and re-evaluated.
This cycle continues until the model achieves satisfactory accuracy.
This paper makes the following {\bf contributions}.
\vspace*{-2pt}
\begin{itemize}
    \item A novel mining method that directly extracts specification flows from input traces to represent advanced system-level protocols in complex system designs.
    \item A new refinement algorithm that can improve the accuracy of the inferred models iteratively until a certain level of accuracy is achieved.
    \item An optimization technique based on the \emph{essential causality} concept to enhance the efficiency of the mining process.
\end{itemize}
\vspace*{-2pt}
Extensive experimental evaluation shows that this method is able to infer models with very high accuracies for diverse traces including synthetic traces and traces generated by executing more realistic system models developed in GEM5~\cite{gem5}.

\begin{figure}[tb]
\begin{center}
\begin{tabular}{p{1.4in}p{1.2in}}
\begin{minipage}{1.2in}
\centering
\includegraphics[width=1.2in]{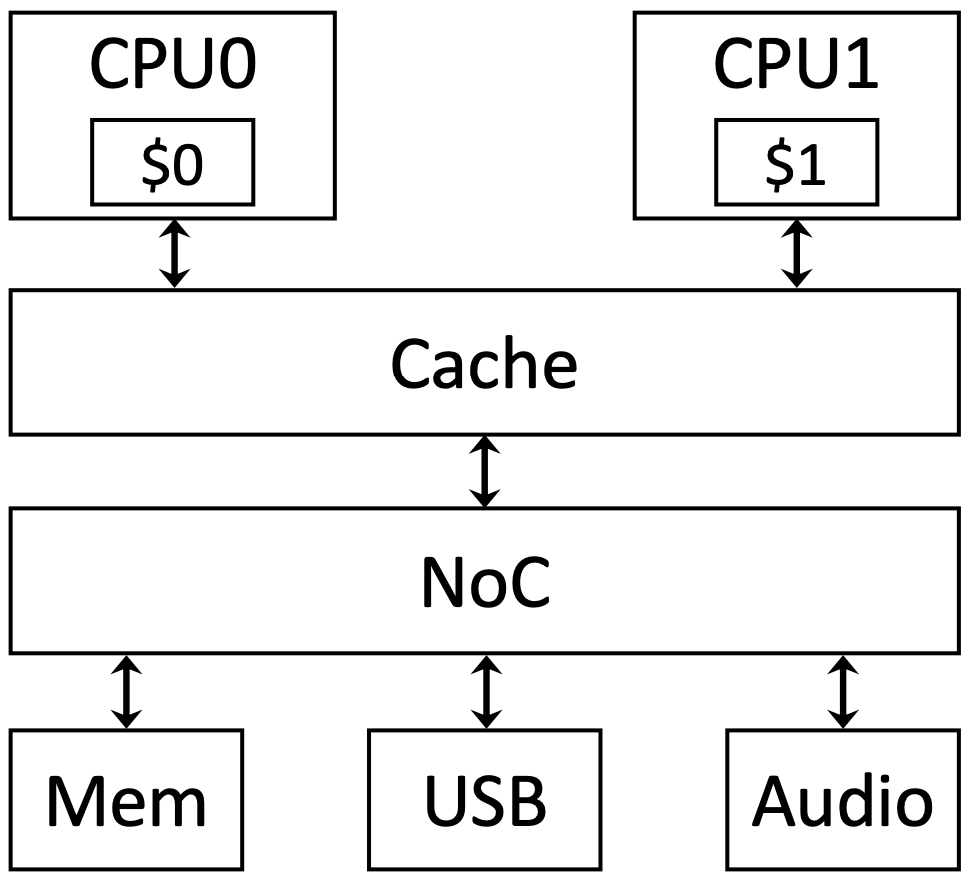}
\end{minipage}
& 
\begin{minipage}{1.2in}
\centering
\includegraphics[width=1.2in]{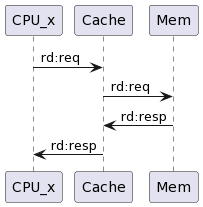}
\end{minipage}
\\
\centering (a) & \centering (b)
\end{tabular}
\caption{ (a) A simplified SoC architecture example, (b) Message sequence diagram for the CPU downstream read flows. This diagram is parameterized with $x$ being $0$ or $1$. Only communications involving CPU, Cache, and Mem blocks are included for a clear illustration.}
\vspace*{-10pt}
\label{fig:soc-ex}
\end{center}
\vspace*{-15pt}
\end{figure}

This paper is organized as follows. Section II reviews some of the previous related works in specification mining and addresses the issues of each one. In section III necessary backgrounds are provided. Sections IV and V represent the proposed method and the proposed optimization methods. In section VI and VII experimental results and some practical use cases are explained. And finally, section VIII provides the conclusion and future works.
\section{Related Works}
Specification mining aims to derive patterns from various artifacts. The \textit{Synoptic} model-based approach extracts invariants from sequential execution logs, recording concurrency partially, and generates a Finite State Machine (FSM) consistent with these invariants \cite{Beschastnikh:2011:LEI:2025113.2025151}.
Another tool, \textit{Perracotta}, analyzes logs of software execution and mines temporal API rules. It employs a chaining technique to identify long sequential patterns \cite{Yang:2006}.
The study known as \textit{BaySpec} extracts LTL (Linear Temporal Logic) formulas from Bayesian networks that are trained using software traces. This approach requires traces to be clearly partitioned with respect to different functions \cite{Mrowca:2019:LTS:3316781.3317847}.
To extract information from hardware traces, the techniques described in \cite{Li2010DAC,Hertz:2013:tcad,Danese:2015:vlsi-soc,Danese:2015:date,Danese:2017:dac} focus on extracting assertions. 
These approaches mine assertions either from gate-level representations \cite{Li2010DAC} or RTL (Register-Transfer Level) models \cite{Chang:2010:aspdac,Hertz:2013:tcad,Danese:2015:vlsi-soc,Danese:2015:date,Danese:2017:dac}.
The research discussed in \cite{Liu:2013} outlines a method for mining assertions by utilizing episode mining from simulation traces of transaction-level models.

Many of these approaches have limitations in detecting longer patterns, making them unable to identify intricate communication patterns that involve multiple components. In contrast, a recent study in~\cite{Ahmed:mine-msg-flow:2021} addresses a comparable problem, but it focuses solely on mining sequential patterns.
The study in \cite{Noc_fabrics} presents a method for validating SoC security properties, focusing on communication fabrics. It involves constructing Control Flow Graphs (CFGs) for each Intellectual Property (IP) component and analyzing the interconnections between these CFGs to assess security risks. The research underscores the importance of system-level interactions and identifies communication fabrics as key areas for security validation. Understanding the communication model at the fabric level is crucial for various validation processes.

Our research focuses on model synthesis methods, aiming to create models from system execution traces that faithfully represent input traces. Heule et al. \cite{Heule:2013} improved the deterministic finite automata inference, initially proposed by Lang \cite{Lang:1998:EDSM}, by treating it as a graph coloring problem and using a Boolean satisfiability solver. \emph{Trace2Model}, a recent development \emph{Trace2Model}, learns non-deterministic finite automata models from software traces using C bounded model checking. Similar research is also noted in \cite{Ulyantsev:2011}. These studies together explore diverse model synthesis techniques, emphasizing accurate system behavior representation from execution traces

The aforementioned methods fail to account for the concurrent nature of communication traces in SoC designs. Instead, they heavily rely on identifying temporal dependencies from the traces to infer models. However, it is important to note that temporal dependencies do not always align with the actual dependencies that our research seeks to uncover.
The \emph{model synthesis} method in~\cite{modelSynthesis}, addresses the problem of the previous approaches and uses a constraint-solving approach to infer finite automata models from SoC communication traces. 
However, it aims to infer models of minimized sizes, not considering the accuracy of the inferred models with respect to the input traces. This approach, assuming that each message leads to at most one subsequent message, is restrictive for more complex communication scenarios.


The \emph{AutoModel} approach~\cite{OurTCAD}, like the \emph{model synthesis} method in \cite{modelSynthesis}, aims to infer abstract models from SoC communication traces. However, it uniquely focuses on inferring models based on the \emph{acceptance ratio} evaluation metric rather than the model size, which enhances the accuracy of the models compared to \emph{model synthesis}. 
Despite this advancement, the core methodology of \emph{AutoModel}, where the final FSA model is composed of certain binary patterns, still carries a significant risk of including invalid message flows.
In this paper, we propose a novel message flow mining method to address the previously mentioned shortcomings.

\section{Background}
\label{sec:backgroun}

\begin{figure}[tb]
\begin{center}
\centering
\begin{tabular}{p{1in}p{1.4in}}
\begin{minipage}{1in}
\centering
\includegraphics[width=.8in]{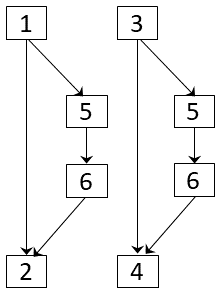}
\end{minipage}
&
\begin{minipage}{1in}
\centering
\begin{small}
\begin{verbatim}
1 (cpu0:cache:rd:req)
2 (cache:cpu0:rd:resp)
3 (cpu1:cache:rd:req)
4 (cache:cpu1:rd:resp)
5 (cache:mem:rd:req)
6 (mem:cache:rd:resp)
\end{verbatim}
\end{small}
\end{minipage}

\\ 
\centering
(a)
& 
\;\;\;\;\;\;\;\;\;(b)
\\
\multicolumn{2}{c}{\begin{small}\texttt{Initial messages = \{1,3\}}\end{small}}
\\
\multicolumn{2}{c}{\begin{small}\texttt{Terminal messages = \{2,4\}}\end{small}}
\\
\multicolumn{2}{c}{\centering (c)} 
\end{tabular}
\caption{(a) Graph representation of CPU downstream flows given in Fig.~\ref{fig:soc-ex}(b) where the nodes are labeled with messages as defined given (b).
(c) gives the flows' initial and terminal messages.
}
\label{fig:flow-ex}
\end{center}
\vspace*{-15pt}
\end{figure}

This section provides the necessary background for the proposed method similar to that in~\cite{OurTCAD}. 
Message flows are a formalism to represent communication protocols across multiple components in a system.
Fig.~\ref{fig:soc-ex}(b) depicts a message flow example for a downstream memory read, focusing on interactions between CPU, Cache, and memory while excluding cache coherence. This paper represents these flows as directed acyclic graphs, as seen in Fig.~\ref{fig:flow-ex}.
Each message, as exemplified in Fig.~\ref{fig:flow-ex}(b), is structured as a quadruple $({\tt src:dest:cmd:type})$, indicating the message's originating and receiving components, the operation at the destination, and whether it is a request or a response respectively.
In Fig.\ref{fig:flow-ex}(b), the message $({\tt cpu0:cache:rd:req})$ is a read request from ${\tt cpu0}$ to ${\tt cache}$. Message flows begin with an initial message to start a flow instance and end with terminal messages signaling completion, encompassing various sequences for different execution scenarios.
Fig.\ref{fig:flow-ex}(a) shows two execution methods for each CPU's memory read flow. The flow from ${\tt cpu0}$ includes sequences (1, 2) for a cache hit and (1, 5, 6, 2) for a cache miss.


Message flows can execute concurrently via \emph{interleaving}, creating traces $\rho = (m_0, m_1, \ldots, m_n)$, with each $m_i$ being a message. In trace $\rho$, if $i < j$, then the ordering is denoted as $m_i <_\rho m_j$. For example, in Fig.~\ref{fig:flow-ex}, if ${\tt CPU0}$ and ${\tt CPU1}$ run their memory read flows three and two times respectively, one possible execution trace is
\begin{equation}
\label{eq:ex-trace}
(3,4,1,1,5,6,2,5,6,2,1,2,3,4).
\end{equation} 

Fig.~\ref{fig:flow-ex}(a) shows that a message flow represents sequences of causality relations among messages, a subset of the broader concept of \emph{structural causality}. This is based on the idea that each message in a system trace is a component's output responding to a preceding input message.
\begin{definition}
\label{def:causal}
Message $m_j$ is {causal} to  $m_i$, denoted as $\mathit{causal}(m_i, m_j)$, if $m_i{\tt .dest} = m_j{\tt .src}.$
\end{definition}

The causality in flow specifications described above is \emph{structural}, distinct from functional or true causality, which includes but is not limited to structural causality. Message flow mining aims to discern all true causality relations from structural ones using system execution traces.

\section{Method}
\label{sec:method}

Algorithm~\ref{algo:overall-algorithm} outlines the method used in this paper, requiring a set of traces, a message definition file, and an accuracy threshold. It generates message flows from the traces that meet the accuracy threshold. It involves 4 main steps: (1) constructing a causality graph from the traces, (2) creating a message flow model by selecting sequences from the graph, (3) evaluating the model's accuracy against the traces, and (4) refining the model if its accuracy is inadequate.
The following subsections give detailed explanations for each of these steps.

\begin{algorithm}[tb]
  \SetKwInOut{Input}{input}
  \SetKwInOut{Output}{output}
\caption{\textbf{Proposed Method}}
\label{algo:overall-algorithm}
\IncMargin{1.5em}
  \Input{A set of traces $T$, Message definition File $F$, Accuracy threshold $\mathcal{A}$, Pruning threshold $\theta$}
  \Output{Message flow model mined from $T$}
  $CG = {\tt ConstructCausalityGraph}(T, F)$\;
  $\mathit{Model}, PG = {\tt ModelSelection}(CG, \theta)$\;
  $\mathit{AR}, \mathit{UM}, \mathit{UE} = {\tt ModelEvaluation}(T, Model)$\; 
  \If{$\mathit{AR} \geq \mathcal{A}$}
  {
    \Return $\mathit{Model}$\;
  }
  \Else
  {
    \Return ${\tt ModelRefinement}(Model, T, \mathit{UM}, \mathit{UE}, \mathit{PG}, \mathcal{A})$\;
  }
\DecMargin{1.5em}
\end{algorithm}
\setlength{\textfloatsep}{0pt}

\subsection{Causality graph Construction}
\label{subsec:cg}

The key to message flow mining is to discern true causalities from structural ones among messages in traces. The initial step in our method is building causality graphs to encapsulate all structural causalities between messages.
\subsubsection{Construction}
\label{subsubsec:construction}

To build causality graphs, the first step is extracting unique messages from input traces $T$, where a unique message differs in at least one attribute from those already collected. We assume that users identify the initial and terminal messages of interest. Let $M$ be the set of all unique messages from $T$, with initial and terminal messages labeled as $i$ and $t$, respectively, where $i \subset M$ and $t \subset M$.

A causality graph is a directed acyclic graph (DAG) with multiple roots and terminals. Each node is labeled with a unique message from set $M$, roots with initial messages from $i$, and terminals with terminal messages from $t$. Edges denote structural causality relations. For each initial message $m_i$ in $i$, a causality graph is built by adding a root for $m_i$, followed by nodes for all messages $m$ where $causal(m_i, m)$ is true, proceeding recursively to terminal messages. Edges forming cycles are omitted in this process.

\begin{figure}
    \centering
    \begin{tabular}{ccc}
        \begin{minipage}{1.1in}
            \includegraphics[width=1.1in]{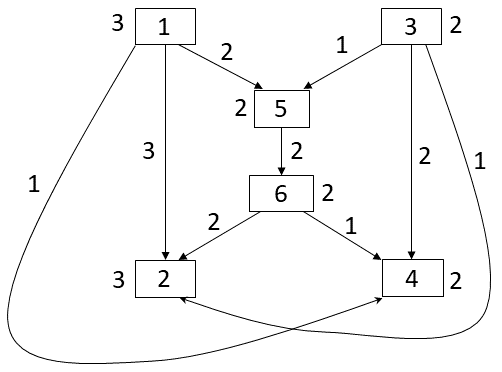}
        \end{minipage}
         &
         \begin{minipage}{1.2in}
             \includegraphics[width=1.2in]{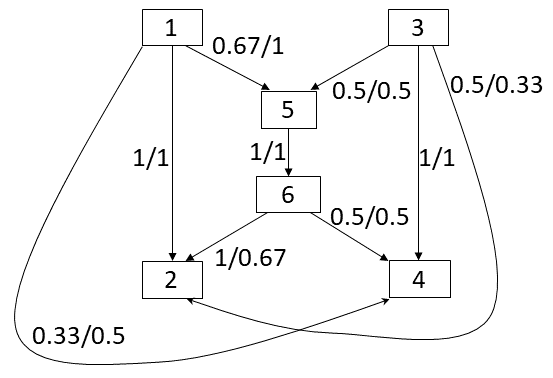}
        \end{minipage}
         &
         \begin{minipage}{0.8in}
             \includegraphics[width=0.8in]{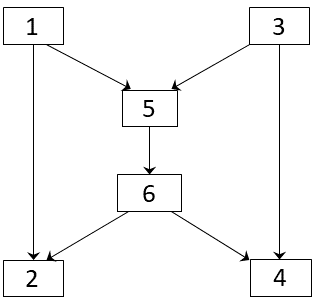}
        \end{minipage}
        \\
         \centering (a) & \centering (b) & \centering (c)
    \end{tabular}
    \caption{Causality graph from trace~(\ref{eq:ex-trace}): (a) nodes and edges with support information, (b) edges marked with forward/backward confidences, (c)  pruned version of the graph.}
    \label{fig:cg-supp-conf-ex}
\end{figure}

\subsubsection{Trace Statistics Collection}
\label{subsubsec:suppotingInfo}

Once the causality graph is built, the next step is to identify message sequences from initial to terminal nodes. In this paper, the terms 'message sequence' and 'path' are used interchangeably. The causality graph, based on input traces, contains all potential paths from initial to terminal nodes. Finding all paths linearly is impractical due to their vast number. To address this, we use trace statistics derived from analyzing the input trace. Definitions~\ref{def:nodeSupport} and \ref{def:edgeSupport} specify node support and edge support respectively.

\begin{definition}
\label{def:nodeSupport}
$NodeSupport(m)$ is the count of instances of a message $m$ in the input trace. 
\end{definition}
\begin{definition}
\label{def:edgeSupport}
$EdgeSupport(h, t)$ for an edge from nodes $h$ to $t$, labeled with messages $m_h$ and $m_t$, is the count of $m_t$ instances each paired with an $m_h$ instance, where $m_h <_\rho m_t$ in the trace and $m_h$ is not matched with other $m_t$ instances.
\end{definition}

After determining node and edge supports, edge confidence measures are defined below. These measures are used to show the causality strength between messages.

\begin{definition}
\label{def:forwardConfidence}
Forward-confidence of edge $h \to t$:
\begin{equation}
\label{eq:forwardConfidence}
ForwardConfidence(h, t)=\frac{EdgeSupport(h, t)}{NodeSupport(h)}
\end{equation} 
\end{definition}

\begin{definition}
\label{def:backwardConfidence}
Backward-confidence of edge $h \to t$:
\begin{equation}
\label{eq:backwardConfidence}
BackrwardConfidence(h, t)=\frac{EdgeSupport(h, t)}{NodeSupport(t)}
\end{equation} 
\end{definition}

Given a set of traces $T$,  the final forward and backward confidences for $T$ are the averages of forward and backward confidences computed for each individual trace $\rho \in T$.

Fig.~\ref{fig:cg-supp-conf-ex} displays the causality graph for trace~(\ref{eq:ex-trace}), where Fig.~\ref{fig:cg-supp-conf-ex}(a) shows node and edge supports, and Fig.~\ref{fig:cg-supp-conf-ex}(b) illustrates forward and backward confidences. This information aids in model selection in the next step.

\subsection{Model Selection}
\label{subsec:modelSelection}

In this work, a \emph{model} comprises message sequences as paths in a causality graph. We first prune the graph by excluding edges with low confidence, then create a base model by selecting paths in the pruned graph that meet specific criteria.

\subsubsection{Pruning the causality graph}
\label{subsubsec:pruning}

Once the causality graph and trace statistics are gathered, we we utilize these statistics to prune the causality graph. 
Edges with higher forward and backward confidence are more likely valid, so those with lower confidence are removed, resulting in a pruned causality graph.

For a better understanding consider the causality graph in Fig.~\ref{fig:cg-supp-conf-ex}.
From trace~(\ref{eq:ex-trace}), all edges except $(1,4)$ and $(3,2)$ have high combined confidences.
Therefore, the edges with lower confidence are deemed likely invalid and are removed from the causality graph. 
Edges like $(3,5)$ and $(6,4)$ with border line confidences might be excluded from the final model. 
The model refinement process using this data is explained in detail later. 
The pruned graph is shown in Fig.~\ref{fig:cg-supp-conf-ex}(c).



It is practical to focus on longer message sequences because they illustrate more complex communication scenarios that involve multiple components, making them significant for validation purposes. 
However, causality graphs might include excessively long message sequences that could be invalid. 
By managing the length of these sequences, it becomes possible to filter out numerous invalid sequences, thereby simplifying the process of model selection.

\begin{algorithm}[tb]
  \SetKwInOut{Input}{input}
  \SetKwInOut{Output}{output}
\caption{\textbf{ModelSelection}}
\label{algo:pathSelection-algorithm}
\IncMargin{1.5em}
  \Input{Causality Graph $CG$} 
  \Output{Model $\mathcal{M}$, Pruned Causality Graph $PG$}
    $PG = Prune Causality Graph(CG)$\; 
    $\mathcal{M} = \emptyset$\;
    Let $\mathit{UM}$ be all messages in $PG$\;
    \While{$\mathit{UM} \neq \emptyset$}
    {
        \ForEach{$initialMessage$ \mbox{in} $PG$}
        {
            Select the longest path $LP$ starting from $initialMessage$ that covers most messages in $UM$\;
            Add $LP$ to $\mathcal{M}$\;
            Let $CM$ be the messages covered by $\mathcal{M}$\;
        $UM = UM - CM$\;
        }
    }
    \Return $\mathcal{M}$, $PG$\;
\DecMargin{1.5em}
\end{algorithm}
\setlength{\textfloatsep}{0pt}

\subsubsection{Selecting the base model}
\label{subsubsec:baseModelSelection}

Model selection aims to find a minimal model meeting the \emph{coverage requirement}, ensuring every message in the causality graph is represented. Following Occam's razor \cite{OccamsRazor}, simpler models offer better generalizability and are easier for users to understand. Hence, the objective is to choose the smallest model satisfying the coverage requirement. In this work, a model's size is determined by the number of message sequences from the pruned causality graph in the inferred model.

To generate the base model, we categorize message sequences in the pruned causality graph by their initial message and length. For each initial message, we add the longest path containing at least one uncovered message to the base model and remove it from the available sequences. Longer sequences are preferred as they represent more complex communication scenarios. This process is repeated until all messages are covered. Details of this path selection algorithm are in Algorithm~\ref{algo:pathSelection-algorithm}.

\subsection{Model Evaluation}

Previous research often evaluated model quality by size \cite{modelSynthesis}, but this does not reliably indicate their effectiveness in explaining input traces. 
To tackle this, our paper uses the \emph{acceptance ratio}, a fractional value measuring the proportion of messages in trace set $T$ accepted by model $M$ relative to the total length of traces in $T$. 
This ratio serves as an indicator of a model's accuracy.

The evaluation algorithm transforms model paths into a finite state automata (FSA), $\mathcal{M} = \{Q, q_0, \Sigma, F, \Delta\}$, consisting of a finite set of states $Q$ with $q0$ as the initial state, a set of symbols $\Sigma$, a subset $F$ of $Q$ for accepting states, and a transition function $\Delta$ mapping from $Q\times\Sigma$ to $Q$.

The evaluation method iterates over all messages in the input trace. 
It starts a new FSA flow instance with each initial message. 
For non-initial messages, it attempts to fit them into active flow instances; if unsuccessful, they are added to unaccepted messages. 
Algorithm~\ref{algo:evaluation-algorithm} represents the evaluation method. 
Unused edges during evaluation and the count of unaccepted messages are returned to be used in the model refinement phase.
\vspace*{-5pt}

\begin{algorithm}[tb]
  \SetKwInOut{Input}{input}
  \SetKwInOut{Output}{output}
\caption{\textbf{ModelEvaluation}}
\label{algo:evaluation-algorithm}
\IncMargin{1.5em}
  \Input{A set of trace $T$, Generated model $\mathcal{M}$}
  \Output{Acceptance Ratio $AR$, Unused Messages $UM$, Unused Edges $UE$}
    Convert $\mathcal{M}$ to an FSA ${M} = \{Q, q_0, \Sigma, F, \Delta\}$\;
    $UnAccepted = \emptyset$\;
    $SumAR = 0$\;
    $X = \emptyset$\;
    $UnusedEdges = \Delta$\;
    \ForEach{$\rho$ in $T$}{
        $Accepted = \emptyset$\;
        \ForEach{$i \in [0, |\rho|]$}{
            $m \gets \rho[i]$\;
            \If{$m$ is a start message} {
                find $q_1$ s.t. $\Delta(q_0, m, q_1)$ holds\;
                Create and add a model instance $({M}, q_1)$ to $X$\;
                Add $m$ to $Accepted$\;
            }
            \ElseIf{$\exists ({M}, q) \in X,~\Delta(q, m, q')$ holds}{
                $X \gets X \cup \{({M}, q')\} - ({M}, q)$\;
                Add $m$ \mbox{to} $Accepted$\;
                Remove $\Delta$ from $UnusedEdges$ if present\;
            }
            \Else{
                Add $m$ \mbox{to} $UnAccepted$\
            }
        }
        $SumAR = SumAR + |Accepted|/|\rho|$\;
    }
    \Return $AR = SumAR/|T|$, $UnAccepted$, $UnusedEdges$\;
\DecMargin{1.5em}
\end{algorithm}

\subsection{Model Refinement}
\label{sec:modelRefinement}
\vspace*{-2pt}

In this phase, our goal is to improve the base model to attain a higher acceptance ratio. We assign scores to paths in the pruned causality graph, and based on these scores and evaluation results, we modify the model by adding or removing paths to increase the acceptance ratio. Subsequent subsections detail this model refinement process.

\subsubsection{Path scoring}
\label{subsubsec:pathScoring}

As previously noted, each edge in the causality graph has trace statistics like forward and backward confidences. Paths in the model, composed of multiple edges, are assigned \emph{forward score} and \emph{backward score}, calculated as the average of confidences of their edges. Higher scores indicate a greater likelihood of a path being valid. The final score of each path is computed using equation~(\ref{eq:finalScore}).
\vspace*{-10pt}

\begin{equation}
\label{eq:finalScore}
  \begin{aligned}[b]
      PathScore= \frac{(ForwardScore+BackwardScore)}{Path Length}
  \end{aligned}
\end{equation}

\subsubsection{Refinement}
\label{subsubsec:refinement}

The refinement algorithm improves the model using evaluation outputs and path scores. If the acceptance ratio meets the accuracy threshold, the model is finalized. Otherwise, it removes unused paths, adds paths with unaccepted messages and high path scores, and re-evaluates. The process repeats until the acceptance ratio meets the threshold or all paths are tested, as detailed in Algorithm~\ref{algo:refinement-algorithm}.

\begin{algorithm}[tb]
  \SetKwInOut{Input}{input}
  \SetKwInOut{Output}{output}
\caption{\textbf{ModelRefinement}}
\label{algo:refinement-algorithm}
\IncMargin{1.5em}
  \Input{Model $\mathcal{M}$, A set of trace $T$, Unaccepted Messages $UM$, Unaccepted Edges $UE$, Pruned Causality Graph $PG$, Accuracy threshold $\mathcal{A}$}
  \Output{Refined Model $\mathcal{M}$}
    $SortedPaths = ComputeScores(PG)$\;
    $AR = 0$\;
    \Repeat{$\mathit{AR} \geq \mathcal{A}$ or $SortedPaths = \emptyset$}
    {
        \ForEach{$edge$ in  $UE$}
        {
            remove $edge$ from $\mathcal{M}$\;
        }
        $maxUnaccepted = max(UM)$\;
        \ForEach{$path$ in $SortedPaths$}
        {
            \If{$maxUnaccepted$ in $path$}
            {
                Add $path$ to $\mathcal{M}$\;
                Remove $path$ from $SortedPaths$\;
                {\bf break}\;
            }
        }
        $AR, UM, UE = {\tt ModelEvaluation}(T, \mathcal{M})$\; 
    }
    \Return $\mathcal{M}$
\DecMargin{1.5em}
\end{algorithm}


\section{Optimization}
\label{sec:optimization}

In this section, we present some techniques designed to enhance the accuracy of the message-mining method and accelerate the evaluation process.

\subsection{Essential Causality Extraction}
We introduce a method for determining true causal relations based on a key insight: \emph{Every message in a trace, except for the initial ones in each message flow, is caused by an earlier message}. 
Specifically, if a message $m$ in a trace is causally linked to exactly one preceding message $m'$ within the same trace, then the causality from $m'$ to $m$ is classified as \emph{essential}. 
Essential causalities are crucial connection points that must be included in at least one of the message flows.

Algorithm~\ref{algo:essential-extract} outlines the procedure for extracting essential causalities. 
Starting with the second message $m_i$ in a trace, the algorithm scans all messages preceding message $m_j$ one by one. 
It checks whether the causality condition $causal(m_j, m_i)$ is met. 
If confirmed, the pair $(m_j, m_i)$ is temporarily stored in set $E$. 
After evaluating all potential $m_j$ for a given $m_i$, this pair is considered essential if $E$ contains exactly one causality pair $(m_j, m_i)$. 
If $E$ contains multiple pairs, this indicates causality ambiguity for $m_i$, and no causality is established. 
The algorithm processes each trace individually, aggregating essential causalities from all traces into a combined set. 
For instance, considering the given trace~(\ref{eq:ex-trace}), starting from the second message, we can determine that the edge from message 3 to message 4 is essential because there are no other causal messages for message 4 apart from message 3. 
Similarly, as we proceed with the algorithm, messages 1-5, 1-2, and 5-6 are also identified as essential for the same reason.

After identifying the essential causalities, we integrate this data into our proposed message flow mining method. 
Initially, during the pruning of the causality graph, edges marked as essential are preserved from pruning. 
Next, in the model selection phase, when the algorithm seeks the longest paths from each initial message to cover the maximum number of uncovered messages, we modify the algorithm to prefer paths that incorporate at least one essential causality. 
In essence, paths containing essential causalities are prioritized over those that do not. 
Lastly, in the model refinement stage, the number of essential causalities within a path is considered when calculating the path score, alongside the average of forward and backward confidences. 
As mentioned before, paths with more essential causalities are more likely to be valid, thus, this adjustment means that paths with a higher count of essential causalities are more likely to be selected for the final model compared to those with fewer essential causalities.


\begin{algorithm}[tb]
  \SetKwInOut{Input}{input}
  \SetKwInOut{Output}{output}
\caption{\textbf{Essential Causality Extraction}}
\label{algo:essential-extract}
\IncMargin{1.5em}
  \Input{A set of trace $T$}
  \Output{A set of essential causalities $EC$}
  $EC \gets \emptyset$\;
  \ForEach{$\rho$ in $T$}
  {
      $F \gets \emptyset$\;
      \ForEach{$i$ \mbox{in} $[1, |\rho|]$}
      {
        $m_i \gets \rho[i]$\;
        $E \gets \emptyset$\;
        \ForEach{$j$ in $[0, i-1]$ }
        {
            $m_j \gets \rho[j]$\;
            \If{$m_j\notin F \wedge$ $causal(m_j, m_i)$  $\wedge$ $(m_j, m_i) \notin E$}
                {Add $(m_j, m_i)$ to $E$;}
        }
        \If{$|E| = 1$}{
            $EC \gets EC \cup E$\;
            Add $m_j$ to $F$\;
        }
      }
    }
  return $EC$\;
\DecMargin{1.5em}
\end{algorithm}

    
\subsection{Essential Message Flow removal}

Many of the message sequences found in trace files, particularly those from GEM5 traces \cite{gem5}, consist of brief message sequences. 
Sequences that contain essential causalities—originating from an initial message and concluding with a terminal message—and where the related messages appear consecutively in the trace file can be immediately removed from the trace file and marked as accepted. 

Given that these messages appear consecutively, starting with an initial message and concluding with the corresponding terminal message, there is a high likelihood that they form a valid message sequence.

Eliminating these messages serves a dual purpose: firstly, it reduces the length of the trace file, which in turn speeds up the evaluation process; secondly, it allows for a focused effort on identifying longer and more intricate message sequences. We call these message flows \emph{Essential Message Flows (EMF)}.

Therefore, prior to initiating the evaluation process, the modified algorithm will cycle through the input trace file and eliminate all EMFs from the trace. Subsequently, the evaluation function will be applied to the remaining messages in the input trace file. The final acceptance ratio reported by the evaluation function will incorporate the number of messages that were removed during the EMF elimination process, as these messages are all considered to be accepted messages.
\vspace*{-5pt}


\section{Experimental Results}
\label{sec:experimentalResults}
\vspace*{-2pt}

The proposed mining method is tested on synthetic traces and traces from system models in the GEM5 simulator \cite{gem5}. This section covers the experimental results and discussions.
\vspace*{-5pt}

\subsection{Synthetic Traces}
\label{subsec:syntheticTraces}
\vspace*{-2pt}
In preliminary tests, our method is applied to synthetic traces generated from $10$ message flows, mimicking real SoC designs, involving memory operations and accesses by CPUs, caches, and peripherals. We created three trace sets: {\tt small-20}, {\tt large-10}, and {\tt large-20}. {\tt small} and {\tt large} refer to the number of message flows, and the numbers indicate flow execution instances. {\tt small-20} includes only CPU-initiated flows ($4$ out of $10$) executed $20$ times, while {\tt large-10} and {\tt large-20} encompass all flows, executed $10$ and $20$ times respectively.
\begin{figure}
    \centering
    \includegraphics[width = 0.8\linewidth]{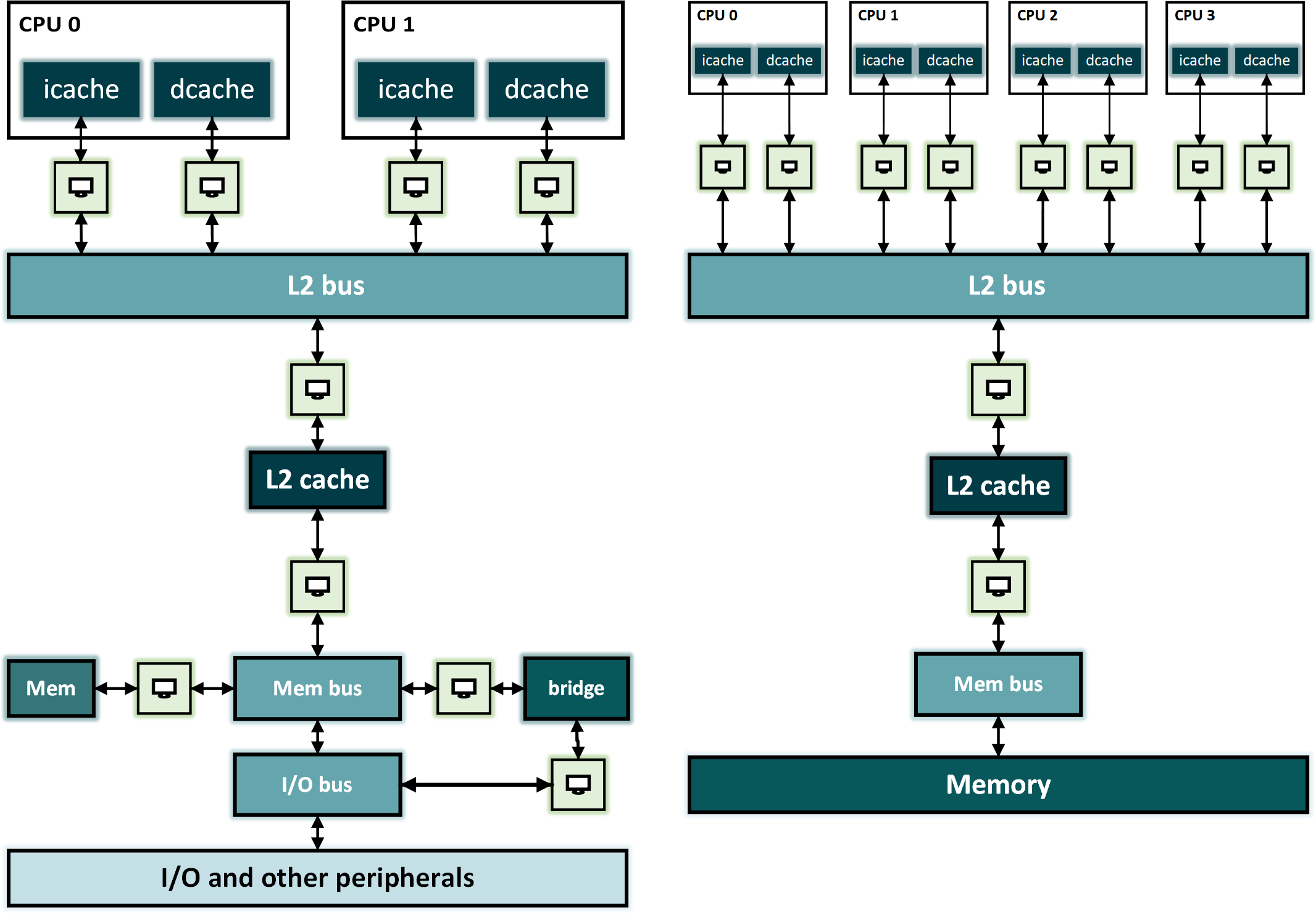}
    \quad \quad\quad \quad \quad (a) \quad \quad \quad \quad \quad \quad \quad \quad \quad (b)
    \caption{(a) GEM5 Full-System (FS) design. (b) GEM5 System Emulation (SE) design. \includegraphics[width=0.15in]{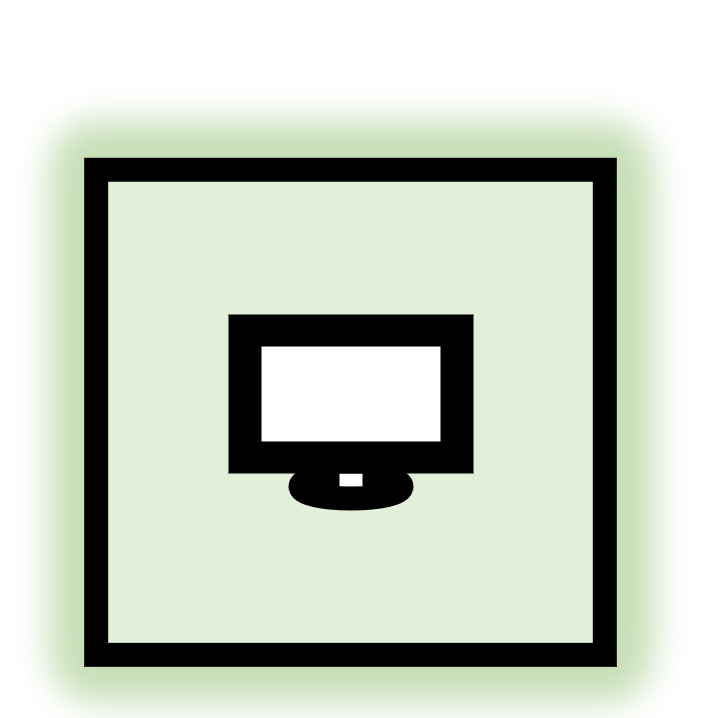}: Communication Monitor}
    \label{fig:gem5}
\end{figure}


\subsection{GEM5 Traces}
\label{subsec:gem5Traces}
\vspace*{-2pt}
For a comprehensive evaluation, we applied our method to GEM5-generated traces. The GEM5 simulator operates in Full System (FS) mode as in Fig.~\ref{fig:gem5}(a) or System-call Emulation (SE) as in Fig.~\ref{fig:gem5}(b). SE mode simulates program-executed system calls and requires static thread-to-core mapping, limiting multi-threaded simulations. FS mode, simulating an operating environment with interrupts, I/O devices, etc., offers more accurate simulations and diverse workloads. The SE setup includes four x86 CPUs with private L1 caches, a shared L2 cache, and 2GB DDR4 memory. The FS setup has two CPUs, DDR3 memory, and additional components like I/O devices.  SE simulation produces trace {\tt threads} from a multithreaded program and trace {\tt snoop} from Peterson’s algorithm. The FS trace comes from Ubuntu 18 Linux OS boot-up.

\subsection{Comparison}
\label{subsec:comparison}

We implemented our method in Python and compared it with the AutoModel method~\cite{OurTCAD}. 
To ensure fairness, we use the acceptance ratio evaluation method proposed in~\cite{OurTCAD} on both AutoModel results and ours for accuracy comparison and compared model sizes based on the number of transitions in the FSAs. 
Results for synthetic and GEM5 traces are shown in Tables~\ref{table:syntheticResults} and \ref{table:gem5Results}, with trace names followed by their message counts. 
In our method no accuracy threshold was set for synthetic and GEM5 snoop traces, instead testing all sequences in the pruned graph. For the larger GEM5 threads, and full system traces, a 97\% and 90\% accuracy threshold was used respectively. The method significantly improved acceptance ratios, by 19.6\% to 33.14\% for synthetic traces and 1.61\% to 19.37\% for GEM5 traces. Notably, AutoModel failed to generate a model for GEM5 threads and full-system traces without code modification and constraint relaxation.

Although the models from our method are larger than those from AutoModel, they extract longer message sequences. Fig.~\ref{fig:pathQuantity-all} compares counts of different length message sequences mined by both methods in GEM5 traces, showcasing our model's ability to reveal complex message sequences.
The proposed method yields significantly more instances of 10-length message sequences than AutoModel. However, for shorter sequences, counts are sometimes lower with our method, implying these cases are encompassed by longer sequences, reflecting our method's higher accuracy compared to AutoModel.

\captionsetup{skip=1pt}
\captionsetup{font=small} 

\begin{table}[!t]
\renewcommand{\arraystretch}{1.3}
\caption{Synthetic Results (The RT is in seconds.)}
\resizebox{\columnwidth}{!}{%
\label{table:syntheticResults}
\centering
\begin{tabular}{|c|c|c|c|c|c|c|c|c|c|c|}
\hline 
\multicolumn{2}{ | c | }{}                                      & \multicolumn{3}{c |}{Large-20 (10900)}         & \multicolumn{3}{c |}{Large-10 (4360)}          & \multicolumn{3}{c |}{Small-20 (3680)}           \\ \cline{3-11}
\multicolumn{2}{ | c | }{}                                      & RT          & Size         & Ratio             & RT          & Size          & Ratio            & RT          & Size          & Ratio             \\ \hline
\multicolumn{2}{ | c | }{Proposed Method}                       & 197         & 134          & \textbf{90.26\%}  & 154         & 132           & \textbf{90.47\%} & 102         & 77            & \textbf{89.57\%}  \\ \hline
\multicolumn{2}{ | c | }{AutoModel \cite{OurTCAD}}              & \textbf{68} & \textbf{103} & 57.12\%           & \textbf{50} & \textbf{107}  & 63.69\%          & \textbf{44} & \textbf{61}   & 69.97\%           \\ \hline
\end{tabular}
}
\end{table}

\begin{table}[!t]
\renewcommand{\arraystretch}{1.3}
\caption{GEM5 Results (The RT is in hours:minutes.)}
\resizebox{\columnwidth}{!}{%
\label{table:gem5Results}
\centering
\begin{tabular}{|c|c|c|c|c|c|c|c|c|c|c|}
\hline
\multicolumn{2}{ | c | }{}                                      & \multicolumn{3}{c |}{Threads (7649395)}         & \multicolumn{3}{c |}{Snoop (485497)}            & \multicolumn{3}{c |}{FullSystem ($8.4\times 10^{9}$)} \\ \cline{3-11}
\multicolumn{2}{ | c | }{}                                      & RT            & Size         & Ratio            & RT            & Size         & Ratio            & RT            & Size         & Ratio                  \\ \hline
\multicolumn{2}{ | c | }{Proposed Method}                       & 2:53          & 234          & \textbf{97.92\%} & 0:42          & 191          & \textbf{92.67\%} & 9:46          & 228          & \textbf{90.04\%}       \\ \hline
\multicolumn{2}{ | c | }{AutoModel \cite{OurTCAD}}              & \textbf{1:26} & \textbf{216} & 96.31\%          & \textbf{0:28} & \textbf{177} & 88.79\%          & \textbf{4:11} & \textbf{214} & 70.67\%                \\ \hline
\end{tabular}
}
\end{table}

\begin{figure*}
    \centering
    \begin{subfigure}[b]{0.34\textwidth}
        \includegraphics[width = \textwidth, left]{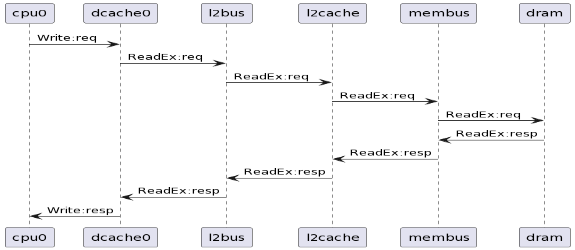}        
        \caption{}
    \end{subfigure}
    \hfill
    \begin{subfigure}[b]{0.34\textwidth}
        \includegraphics[width = \textwidth, right]{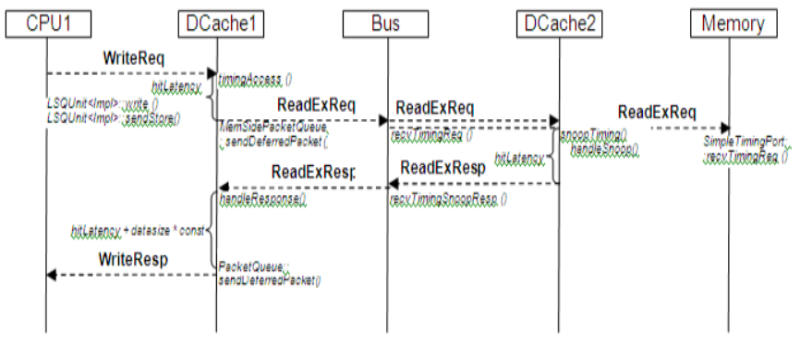}
        \caption{}
    \end{subfigure}
    \hfill
    \begin{subfigure}[b]{0.275\textwidth}
        \includegraphics[width = \textwidth, right]{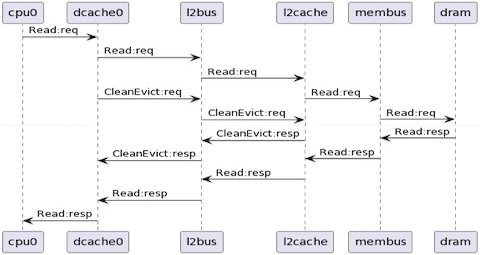}
        \caption{}
    \end{subfigure}
     
    \caption{(a) A flow mined from the GEM5 threads trace. (b) GEM5 documentation's memory write miss sequence. (c) Two flows mined from the GEM5 threads trace depicting a clean eviction scenario, \textit{not in GEM5 documentation}.}
    \label{fig:MixedExample}
\end{figure*}


\begin{figure*}
    \centering
    \begin{subfigure}[b]{0.32\textwidth}
        \includegraphics[width = \textwidth, left]{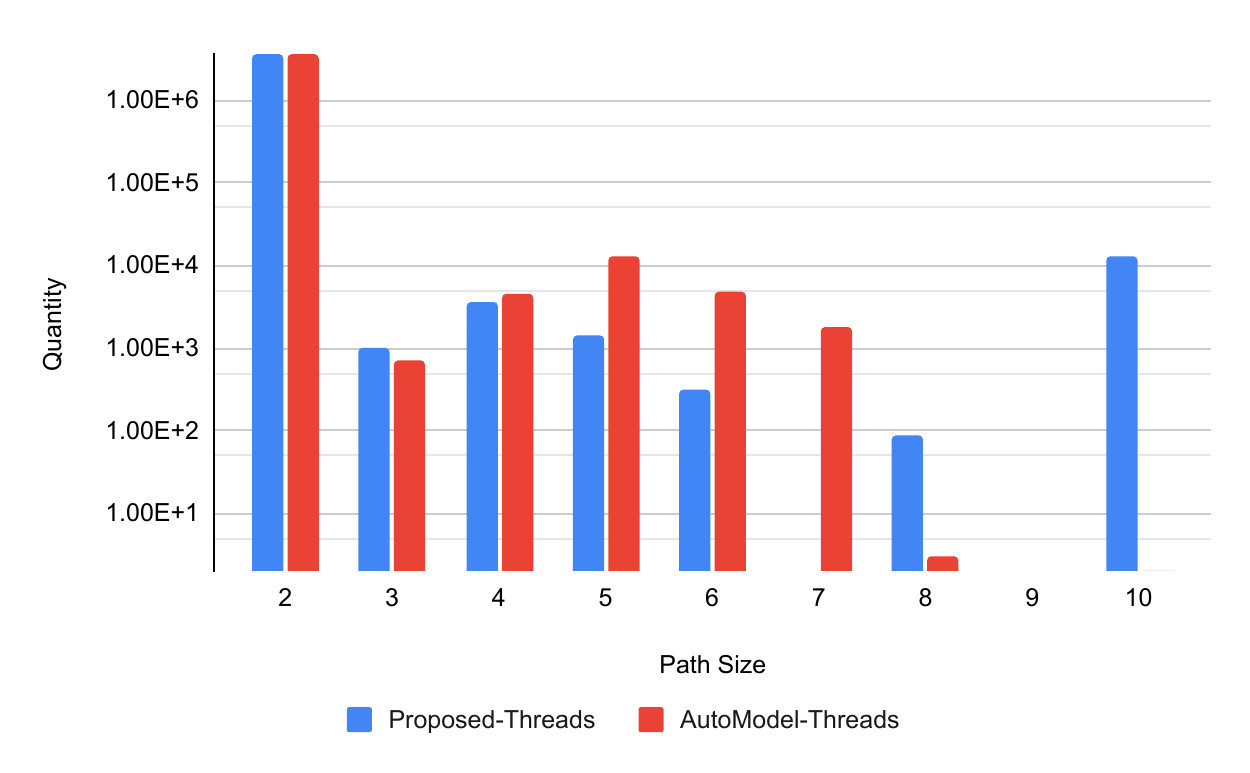}        
        \caption{}
    \end{subfigure}
    \hfill
    \begin{subfigure}[b]{0.32\textwidth}
        \includegraphics[width = \textwidth, right]{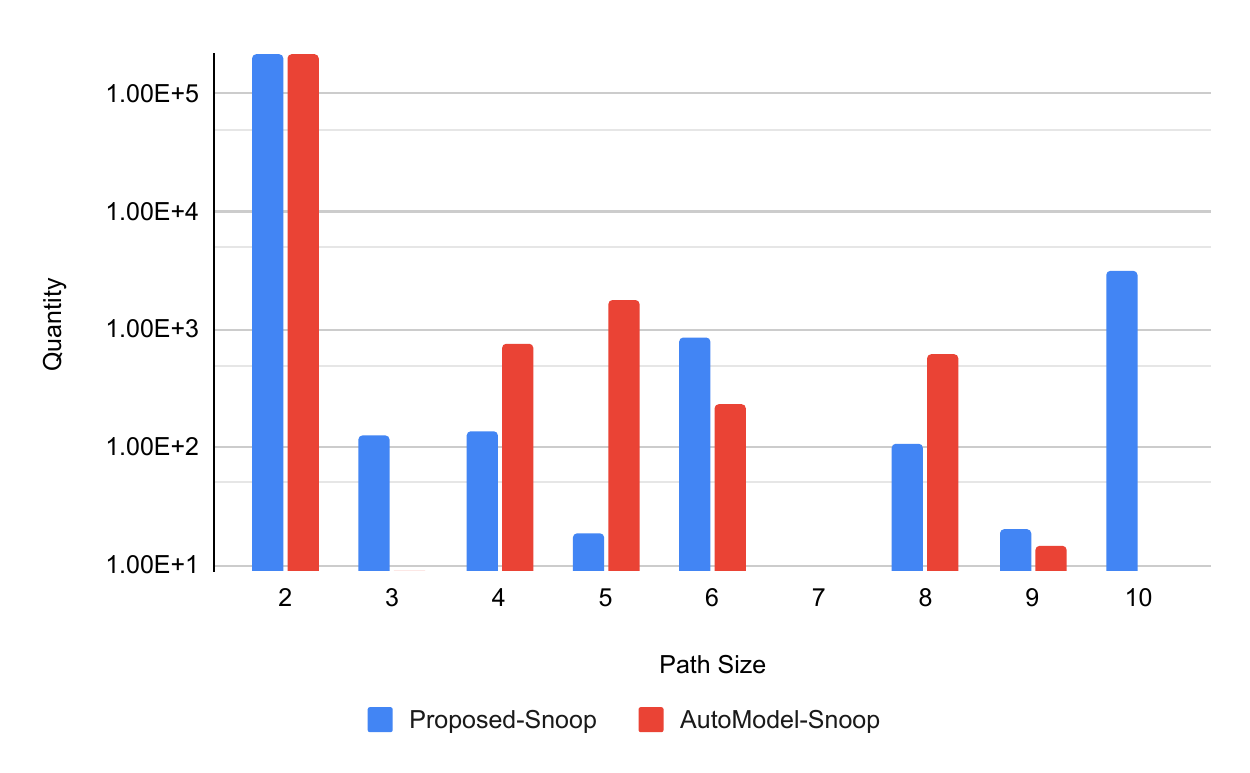}
        \caption{}
    \end{subfigure}
    \hfill
    \begin{subfigure}[b]{0.32\textwidth}
        \includegraphics[width = \textwidth, right]{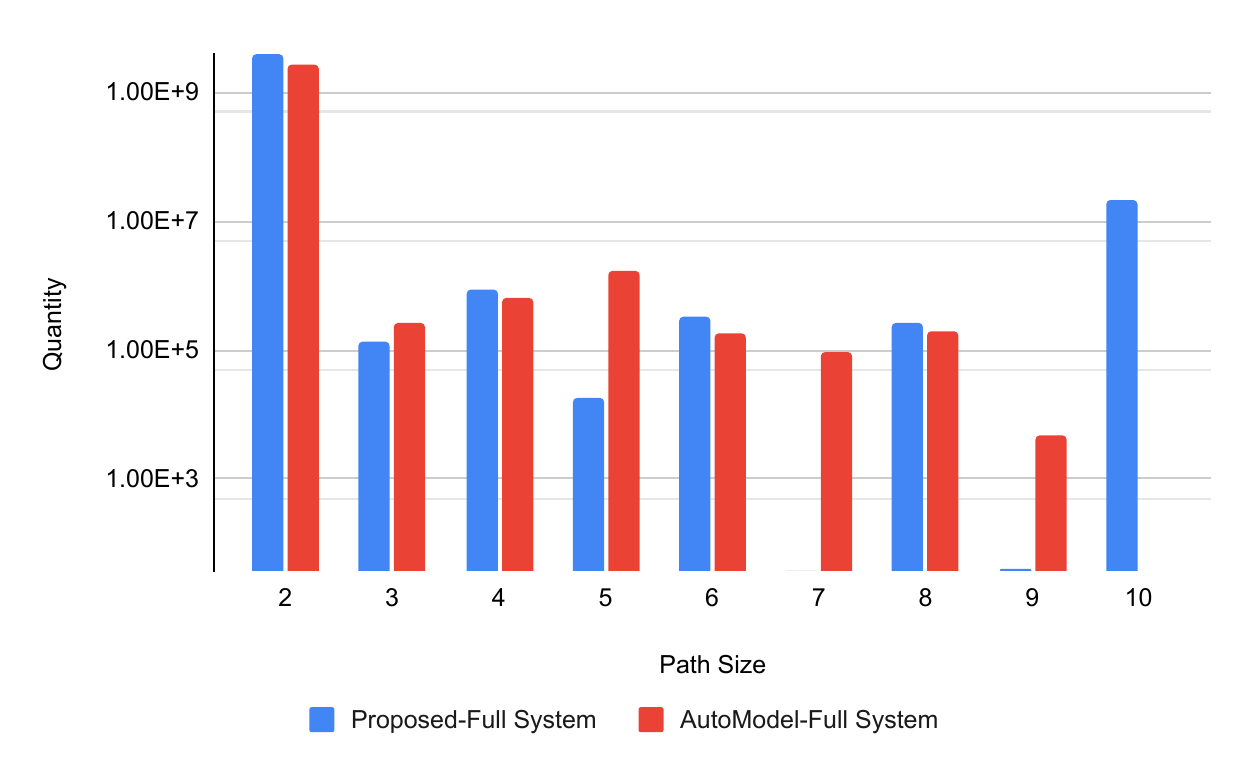}
        \caption{}
    \end{subfigure}
     
    \caption{Counts of instances of the mined message sequences of different lengths using the proposed method and AutoModel \iffalse model synthesis\fi method that are found in the GEM5 (a) {\tt threads}, (b) {\tt snoop}, and (c) {\tt full system} traces.}
    \label{fig:pathQuantity-all}
\vspace*{-15pt}
\end{figure*}

\subsection{Discussions}
\label{subsec:discussion}
There is a trade-off between model accuracy and size, and between accuracy and refinement runtime. Lowering the accuracy threshold speeds up refinement and reduces model size, while a higher threshold increases both runtime and size. The significant runtime of our method for the FS trace is due to the evaluation phase. Following each model refinement, an evaluation is conducted, contributing to the extended runtime. User vision can enhance the model's efficiency. For instance, incorporating specific paths into the model can boost accuracy, thereby decreasing the number of refinement iterations and finally reducing runtime. Based on our experiments, beyond a certain accuracy level, further improvements significantly increase model size and runtime. Thus, user insight is vital to find the optimal accuracy threshold.
 
\section{Pratical uses of the mined message flows}
\label{subsec:usecases}

The proposed message flow mining method holds potential utility across various fields such as verification, debugging, enhancing test coverage, security, etc. 

\subsection{Better Design Understanding}
SoC communication traces are typically very long and unwieldy for human to understand directly. 
Mined message flows provide an abstract representation of the SoC communication behavior, and help to uncover complex dependencies and interactions that might not be immediately evident. 
For example, Fig.~\ref{fig:MixedExample}(a) shows a message sequence with length 10 from the GEM5 threads trace, illustrating a write operation during a cache miss, as inferred by our method. 
Fig.~\ref{fig:MixedExample}(b) shows the corresponding protocol for handling write misses as in the official GEM5 documentation, and it is incomplete in terms of responses from memory to L2 cache {\tt DCache2} as compared with the mined protocol as shown in Fig.~\ref{fig:MixedExample}(a).
This example highlights how the mined message flows offer a deeper understanding of the actual system designs. 
Additionally, Fig.~\ref{fig:MixedExample}(c) shows the mined message flow for memory read and cache clean eviction scenarios.
However, the part in Fig.~\ref{fig:MixedExample}(c) for the cache clean eviction scenario is not included in the GEM5 documentation.
\vspace*{-5pt}

\subsection{Verification Against Design Specifications}
In instances where a golden specification documentation is available, the mined message flows can assist in identifying potential deviations or violations in hardware behavior. 
By comparing the mined message flows from the execution traces with the golden documentation, any discrepancies between the mined message flows and the documented flows can indicate potential deviations in the hardware's behavior from its intended design.
Also, the mined message flows from the execution traces allowing us to identify and address any specification violations quickly. 
This ensures that the hardware functions as intended and meets all design requirements.
\vspace*{-5pt}

\subsection{Efficient Debugging}
In situations where a golden design specification documentation is not available, which is often the case, the mined message flows can significantly aid in identifying, localizing, and determining the root causes of bugs in the system.
\subsubsection{Bug Detection}
Message flow mining helps in detecting unusual patterns or sequences in the communication between hardware components. 
This structured information accelerates the debugging process by narrowing down the potential causes of errors.
Also, essential causalities identified through message flow mining can help prioritize critical bugs that have a higher impact on the system’s functionality.

\subsubsection{Bug Localization}
The method used in \cite{Li2010DAC} involves identifying recurring temporal patterns from correct and erroneous traces, which are then used to detect inconsistencies. 
These distinguishing patterns help localize errors within the system, proving particularly useful for post-silicon debugging by addressing challenges such as limited observability and reproducibility of bugs.

Building upon this concept, our approach to solving the bug localization problem focuses on message flow mining. 
For transient errors that occur under specific conditions, message flow mining helps isolate the problem by comparing message flows from healthy traces (captured during error-free operations) with those from faulty traces (captured during system failures). 
This comparison allows us to identify the location and conditions where the bug occurs. 
For permanent errors, comparing the mined message flows with the design documentation helps in identifying consistent discrepancies, enabling us to locate and fix the root cause of the error.

To illustrate this, we intentionally reconfigured the communication monitor between {\tt cpu0} and {\tt dcache0} from the GEM5 snoop trace configuration file to drop the messages destined for a specific memory address and re-ran the simulation to simulate a system malfunction. 
We then mine message flows from the generated buggy trace to identify the missing parts. 
By comparing the results from the buggy trace file and the untampered healthy trace file, we observe an 11.83\% drop in the acceptance ratio. 
This significant decrease is due to the inability of the mined model to find complete flows for messages where parts are missing. 
Furthermore, comparing the mined message flows from the healthy and buggy trace files, we find that the buggy trace is missing four different message flows that involve the removed communication monitor interface. 
This experiment demonstrates that mined message flows can effectively identify and locate bugs within the system.

\subsubsection{Facilitating Root Cause Analysis}
Message flow mining allows us to trace back from the point of failure to the root cause by following the sequence of message interactions. 
This backward analysis helps in understanding the chain of events leading to the error.
\vspace*{-5pt}

\subsection{Other Use Cases}
\vspace*{-5pt}

\subsubsection{Uncovering Untested Scenarios} 
Message flow mining can reveal areas of the hardware that are not adequately tested by the existing test cases. By identifying missing or infrequent message flows, new test cases can be created to cover these scenarios, ensuring more comprehensive testing.
\subsubsection{Improving Confidence in Verification} 
Thorough test coverage improves the overall confidence in the hardware's reliability and performance, reducing the risk of undetected bugs.
\subsubsection{Monitoring Changes} 
When changes are made to the hardware design, message flow mining can be used to verify that these changes do not introduce new errors. By comparing message flows before and after the changes, we can ensure that the modifications have not negatively impacted the system.
\subsubsection{Real-time Monitoring and Security} 
Continuous monitoring of message flows can help in the real-time detection of anomalies, enabling prompt intervention and reducing the impact of bugs, and also probable security breaches.


\section{Conclusion}
\vspace*{-5pt}
This paper presents a method to accurately infer message flow specifications from complex system traces, outperforming existing methods. 
Our method excels in handling large, intricate traces and provides insights into system protocols and component interactions, aiding in system design, debugging, and efficiency improvements. 
The concept of essential causalities aids in pinpointing critical components of the design, mining more accurate message flows, and accelerating the mining process.
It should be noted that this mining approach isn't entirely automated; it requires user input for enhanced performance, serving as a tool for designers. Future work includes optimizing runtime with trace slicing, enhancing accuracy with techniques like window slicing, and exploring deep learning for identifying valid paths using high-accuracy sample models.
\vspace*{-10pt}

\end{document}